# Attenuation map reconstruction from TOF PET data

*Qingsong Yang, Wenxiang Cong, Ge Wang\**

*Department of Biomedical Engineering, Rensselaer Polytechnic Institute,
Troy, NY 12180, USA*

*\*Ge Wang (ge-wang@ieee.org)*

**Abstract:** To reconstruct a radioactive tracer distribution with positron emission tomography (PET), the background attenuation correction is needed to eliminate image artifacts. Recent research shows that time-of-flight (TOF) PET data determine the attenuation sinogram up to a constant, and its gradient can be computed using an analytic algorithm. In this paper, we study a direct estimation of the sinogram only from TOF PET data. First, the gradient of the attenuation sinogram is estimated using the aforementioned algorithm. Then, a relationship is established to link the differential attenuation sinogram and the underlying attenuation background. Finally, an iterative algorithm is designed to determine the attenuation sinogram accurately and stably. A 2D numerical simulation study is conducted to verify the correctness of our proposed approach.

## 1. Introduction

In positron emission tomography (PET), the attenuation background of the tissue is necessary to reconstruct a radioactive tracer distribution. Generally, this attenuation correction can be estimated from a CT scan in a PET-CT system. However, this CT scan may be inaccurate due to patient motion between the PET and CT scans. Moreover, there are situations where a CT scan is not available due to the radiation dose. PET image reconstructed with an incorrect attenuation map or without this information would suffer from significant attenuation artifacts.

Time-of-flight (TOF) PET was developed decades ago but it only recently came into practice, thanks to the ultrafast electronics and scintillation material. In TOF PET, a pair of photons from an annihilation location is measured in a time-resolving fashion. While it is known that an attenuation correction map cannot be uniquely found only from PET data [1], a recent study demonstrated that TOF-PET data contain important information on attenuation coefficients. In [2], a maximum-a-posterior reconstruction algorithm was proposed to simultaneously reconstruct both radioactive activities and attenuation coefficients in TOF-PET. In [3], the gradient of the attenuation sinogram was proved to be uniquely computable only from TOF PET data. However, up to date there was no scheme proposed to estimate the absolute attenuation background from the gradient of the attenuation sinogram. To find this constant term, a prior knowledge on the attenuation background was suggested in [3].

The main contribution of this paper is to eliminate the above-described constant uncertainty for self-sufficient TOF PET imaging. Our idea was obtained from the field of differential projection imaging where it is feasible to exactly reconstruct an image from derivatives of the involved sinogram using an analytical or iterative algorithms [4-6]. In this work, we demonstrate that TOF-PET data can accurately and stably determine an attenuation sinogram and the attenuation correction map without any specific knowledge on it. The least-square estimation method will be used to estimate derivatives of an attenuation sinogram [3], and an image reconstruction method is designed for TOF PET. The second section describes the methodology in detail. The third section reports a numerical simulation study to verify the formulation and the algorithm. In the fourth section, relevant issues are discussed, and the conclusion is drawn.

## 2. Methodology

### 2.1. TOF-PET Data Model

Let a radioactive tracer distribution be denoted as $f(x, y)$. In 2D conventional PET, the measurement along a line is expressed as

$$m(\theta, s) = p(\theta, s) e^{-g(\theta, s)}, \quad (1)$$

where $p(\theta, s)$ is the generic PET data without attenuation in parallel-beam geometry,

$$p(\theta, s) = \int_{-\infty}^{\infty} f(s\cos\theta - l\sin\theta, s\sin\theta + l\cos\theta) \, dl, \quad (2)$$

and $g(\theta, t)$ is the Radon transform of the attenuation background $\mu(x, y)$,

$$g(\theta, s) = \int_{-\infty}^{\infty} \mu(scos\theta - lsin\theta, ssin\theta + lcos\theta) \, dl. \tag{3}$$

In TOF-PET, due to the limited time resolution the measurement can be modeled as

$$p(\theta, s, t) = \int_{-\infty}^{\infty} f(x, y)\delta(xcos\theta + ysin\theta - l) \, w(t - l) dl, \tag{4}$$

where $w(t)$ is a time profile, which is assumed as a Gaussian function with a standard deviation $\sigma < \infty$,

$$w(t) = \frac{1}{\sqrt{2\pi}\sigma} e^{-t^2/2\sigma^2} \tag{5}$$

Recent research demonstrates that the attenuation sinogram is determined by TOF PET data up to a constant shift, as stated in the following theorem:

**Theorem 1** [2]: *The emission data $m(\theta, s, t)$ determine derivatives of the Radon transform $g(\theta, s)$ over $\theta$ and $s$ if (1) The TOF time profile is a Gaussian function; (2) for each measured line of response (LOR), the TOF data are measured for all $t \in \mathbb{R}$; (3) $f(x, y)$ and $\mu(x, y)$ are non-negative functions with continuous first derivatives and bounded supports; and (4) No LOR is totally attenuated so that $e^{g(\theta,s)} > 0$ for all $\theta$ and $s$.*

Based on the proof of the theorem, an analytical scheme for estimation of the gradient of the attenuation sinogram is given as follows [2]:

$$\frac{\partial g}{\partial s} = -\frac{J_s H_{\theta\theta} - J_\theta H_{s\theta}}{H_{ss} H_{\theta\theta} - H_{s\theta}^2}$$

$$\frac{\partial g}{\partial \theta} = -\frac{J_\theta H_{ss} - J_s H_{s\theta}}{H_{ss} H_{\theta\theta} - H_{s\theta}^2} \tag{6}$$

Where

$$H_{ss} = \int_\tau \left(mt + \sigma^2 \partial_t m\right)^2 dt, \quad H_{s\theta} = \int_\tau m\left(mt + \sigma^2 \partial_t m\right) dt, \quad H_{\theta\theta} = \int_\tau m^2 dt,$$

$$J_s = \int_\tau (D[m])\left(mt + \sigma^2 \partial_t m\right) dt, \quad J_\theta = \int_\tau (D[m]) m \, dt, \tag{7}$$

and the operator $D[\cdot]$ is defined as

$$D[m(\theta,s,t)] = t\frac{\partial m}{\partial s} + \frac{\partial m}{\partial \theta} - s\frac{\partial m}{\partial t} + \sigma^2 \frac{\partial^2 m}{\partial s \partial t} \tag{8}$$

Based on Theorem 1, here we propose to utilize a differentiated backprojection imaging method to uniquely determine the attenuation sinogram from TOF PET data m(θ, s, t) under the conditions of Theorem 1. In fact, using the relationship between the backprojection of differentiated attenuation projection data and the Hilbert transform of the attenuation background, we have [7]

$$H_\theta \mu(x) = -\frac{1}{2\pi} \int_{\theta_0}^{\theta_0+\pi} \frac{\partial g(s,\theta)}{\partial s}\bigg|_{s=x\cdot(\cos\theta,\sin\theta)} d\theta \tag{9}$$

where H is the Hilbert transform along direction $\vec{n} = (-\sin\theta, \cos\theta)$. Eq. (9) can be rewritten as follows:

$$\mu(x) = -\frac{1}{2\pi} H_\theta^{-1}\left(\int_{\theta_0}^{\theta_0+\pi} \frac{\partial g(s,\theta)}{\partial s}\bigg|_{s=x\cdot(\cos\theta,\sin\theta)} d\theta\right) \tag{10}$$

From Eq. (10), the attenuation background can be uniquely determined from the derivatives of an attenuation sinogram. Furthermore, by Theorem 1 the TOF PET data uniquely determine derivatives of the attenuation sinogram. Hence, the TOF PET data $m(\theta, s, t)$ can uniquely determine the attenuation sinogram, and this inversion process is accurate and stable as well. A more efficient analytical reconstruction method for reconstruction of the attenuation background from derivatives of the attenuation sinogram is given in the following subsection.

**2.2. Reconstruction of an Attenuation Background from Gradient Data**

Since Eq. (6) gives the gradient data of an attenuation sinogram, the problem is reduced to reconstruct the attenuation map $\mu(x,y)$ from the gradient data. The classical filtered backprojection (FBP) reconstruction algorithm can be expressed as

$$\mu(x,y) = \mathcal{B}\mathcal{F}^{-1}\big[|\omega|\mathcal{F}[g(\theta,s)]\big] \tag{11}$$

where $\mathcal{B}$ is the backprojection operator, $\mathcal{F}$ is the Fourier transform operator, and $|\omega|$ is the ramp filter. According to the differential property of the Fourier transform, we have

$$\mathcal{F}[\partial_s g] = 2\pi i \omega \mathcal{F}[g(\theta, s)] \qquad (12)$$

Substitute Eq. (12) into Eq. (11), we have

$$\mu(x,y) = \mathcal{B}\mathcal{F}^{-1}\left[|\omega|\frac{1}{2\pi i \omega}\mathcal{F}[\partial_s g]\right] = \mathcal{B}\mathcal{F}^{-1}\left[\frac{\text{sign}(\omega)}{2\pi i}\mathcal{F}[\partial_s g]\right] \qquad (13)$$

Hence, the TOF PET attenuation background can be uniquely reconstructed from the *s* derivatives of the attenuation sinogram using an adapted FBP algorithm with $\text{sign}(\omega)/2\pi i$ as its filter kernel. Similarly, other reconstruction formulas can be constructed from different combinations of available partial derivatives.

## 3. Simulation Results

### 3.1. Experimental Design

To verify our formulation and algorithm, a 2D TOF PET simulation study was performed. The numerical phantom we used had most of the parameters the same as that in [2], shown in Figure 1. The field of view (FOV) was set to 40cm in diameter, and sampled into an images of $384 \times 384$ pixels (pixel size 0.104cm). The real attenuation sinogram was obtained in parallel projections uniformly over an $180^0$ angular range. TOF-PET data were synthesized by convolving the image with the 1D Gaussian profile of a standard deviation $\sigma_t$. For convenience, let $N_s$ denote the number of detectors, $N_\theta$ the number of view angles, and $N_t$ the number of temporal bins. Hence, the simulated TOF data were a tensor of $N_s \times N_\theta \times N_t$ with sampling steps $\Delta_s$ along *s*, $\Delta_\theta$ along $\theta$, and $\Delta_t$ along *t* respectively. In all of the simulation tests, $N_t$ was 128 with the corresponding $\Delta_t$ being translated to 0.3125ns so that the temporal slices could cover the whole image [8].

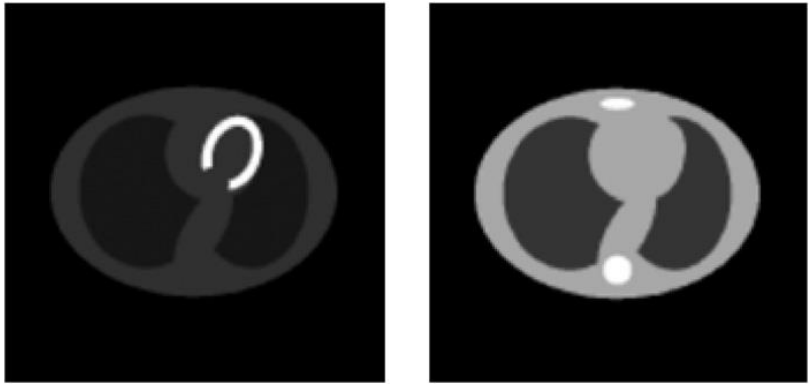

Figure 1. Numerical phantom of a radioactive tracer distribution (left) on an attenuation background (right).

The derivatives of the attenuation sinogram $\partial_s g(\theta, s)$ were approximated using Eq. (6). The derivatives were computed as finite differences, and the integrals over *t* approximated as a Riemann sum. To suppress noise and errors from the finite differences, the TOF PET data were smoothed using two Gaussian kernels along *s* and *θ* with variance $\sigma_s$ and $\sigma_\theta$ respectively. In our experiments, $\sigma_s$ was $1.3\Delta_s$ for $N_s$=128, and $1.8\Delta_s$ for $N_s$=256, and $\sigma_\theta$ was equal to $\Delta_\theta$. It was found that $\sigma_s$ had more impact than $\sigma_\theta$.

Finally, the attenuation map $\mu(x, y)$ was reconstructed using the filtered backprojection method described in the second section. For that purpose, the MATLAB function *iradon()* was modified by replacing the filter with the sign function according to Eq. (13). The size of the reconstructed image was determined by $N_s$. To obtain the attenuation sinogram from the reconstructed image, the forward projection operation was performed after transforming the image to the original size.

First, a noise free simulation test was performed. The temporal resolution was 500ps, and the full-width at half-maximum (FWHM) was 7.5cm and $\sigma_t = \frac{FWHM}{2\sqrt{2\ln 2}} \approx 3.185\ cm$. The TOF PET data were $128 \times 128 \times 128$, which means 128 projections, 128 detectors per projection, and 128 temporal steps. The reconstructed image is in Figure 2. The associated sinogram is in Figure 3. To compare the re-projected sinogram and the true values, three vertical profiles were plotted at columns 32, 64, and 96 respectively in Figure 4.

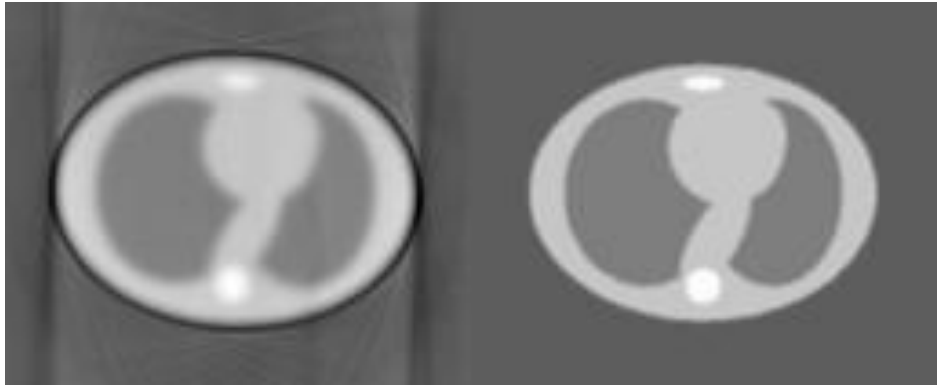

Figure 2. Reconstructed attenuation map (left) and the truth (right). The display window is [-0.08, 0.14].

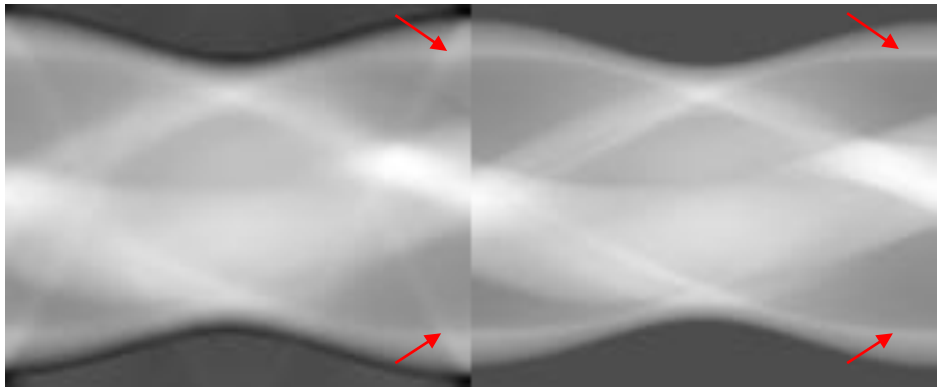

Figure 3. Reprojected sinogram (left) and the truth (right). The display window is [-0.98, 2.26].

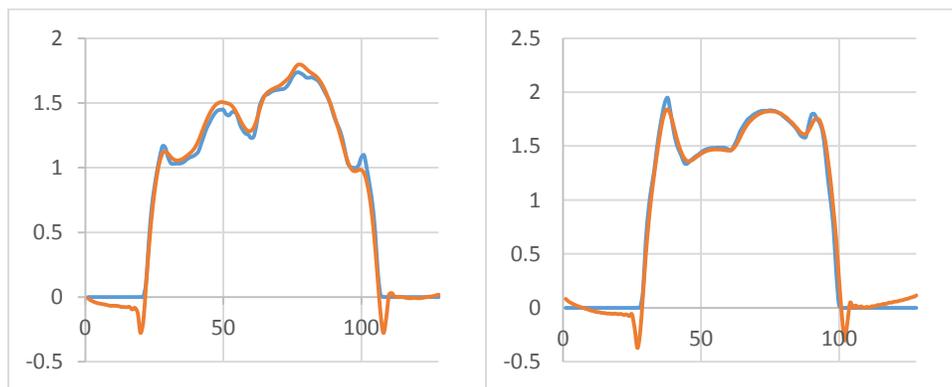

(a)                                              (b)

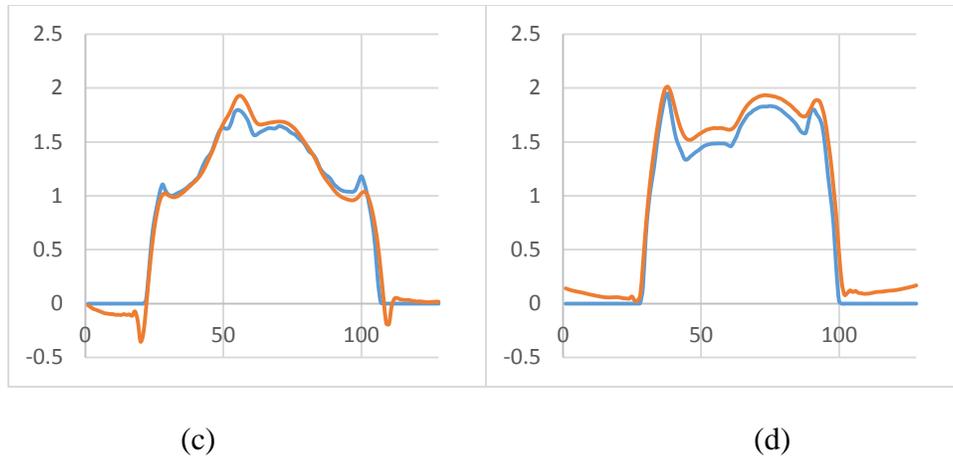

(c)                             (d)

Figure 4. Profiles comparison in the sinogram domain. The red curves show estimated values, while the blue counterparts are true values. In (d), a non-negative correction step was applied to the image.

### 3.2. Image quality assessment

The performance of our algorithm was shown by the images in Figure 3 and the profiles in Figure 4. In the profiles, the estimated values matched the true values in most parts. However, some artifacts were still produced due to numerical errors.

One of the error-prone areas in the image domain was across the boundaries. The values of black pixels were negative, which is impossible in reality. The boundary errors were from the estimated gradient of the attenuation sinogram, as illustrated in Figure 5. Unfortunately, these boundary errors cannot be avoided with the current estimation method [2], and our current reconstruction method used these unreliable data. One easy solution is to set the negative values to be zero. However, this non-negative correction in the image domain would add a positive bias in the sinogram, as shown in Figure 4 (d).

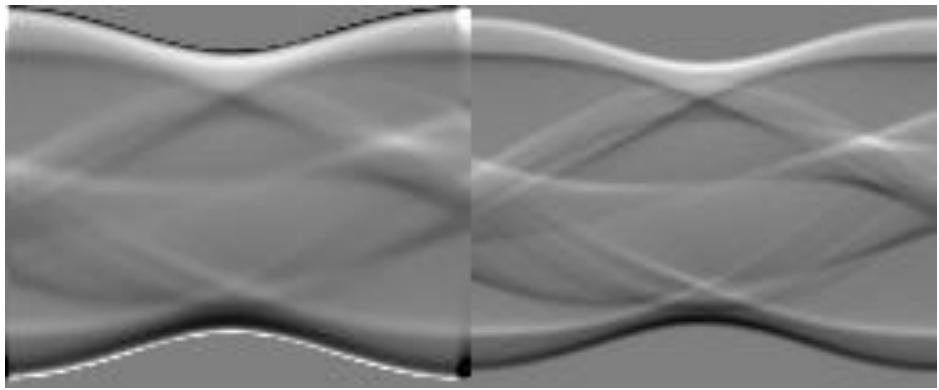

Figure 5. Estimated gradient of the sinogram (left) and the truth (right). The display window is [-1.3, 1.3].

Also, some artifacts are indicated with the red arrows in Figure 3. These artifacts were generated by pixels outside the phantom support. Ideally, the reconstructed pixel values outside the elliptical support should be zero. However, the analytical reconstruction method cannot enforce the phantom support automatically. To eliminate these artifacts, we manually set those pixels that were outside the phantom support to zero and reproduced the forward projections in Figure 6.

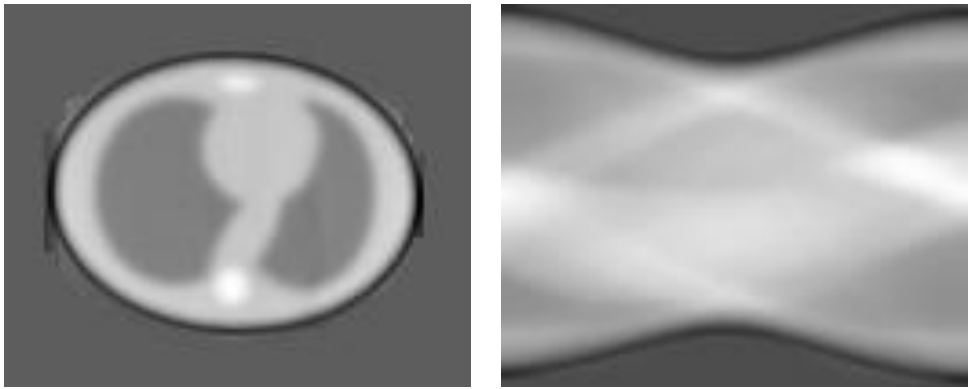

Figure 6. Manually cleaning the pixels outside the attenuating background (left) and resulting in an improved attenuation sinogram (right).

## 4. Discussions and conclusion

Recent advance in TOF PET research has demonstrated that TOF PET data can determine partial derivatives of the attenuation sinogram. In this paper, we have established that the sinogram can be uniquely and stably determined only from TOF PET data, without the constant ambiguity.

In our proposed method, the sinogram is indirectly obtained, and errors are accumulated through the process. The boundary errors in the estimation of the gradients of the attenuation sinogram may result in negative ripples in the sinogram. These artifacts can be effectively corrected using the non-negativity constraint properly.

In our simulation study, to eliminate the artifacts, the attenuation image was manually modified. This method can be used in most cases. However, it is not efficient. The best way is to reconstruct an image accurately. To achieve this goal, further work should focus on two aspects. One is to compute the derivative information more accurately from TOF-PET data. The other is to design a better reconstruction algorithm which can use all available partial derivatives of the attenuation sinogram.